# Microwave whirlpools in a rectangular-waveguide cavity with a thin ferrite disk

E.O. Kamenetskii, Michael Sigalov, and Reuven Shavit

Department of Electrical and Computer Engineering,
Ben Gurion University of the Negev, Beer Sheva, 84105, Israel

July 9, 2006

**Abstract**

We study a three dimensional system of a rectangular-waveguide resonator with an inserted thin ferrite disk. The interplay of reflection and transmission at the disk interfaces together with material gyrotropy effect, gives rise to a rich variety of wave phenomena. We analyze the wave propagation based on full Maxwell-equation numerical solutions of the problem. We show that the power-flow lines of the microwave-cavity field interacting with a ferrite disk, in the proximity of its ferromagnetic resonance, form whirlpool-like electromagnetic vortices. Such vortices are characterized by the dynamical symmetry breaking. The role of ohmic losses in waveguide walls and dielectric and magnetic losses in a disk is a subject of our investigations.

## 1. Introduction

A three-dimensional system of a rectangular-waveguide resonator with an inserted thin ferrite disk is a very attractive object for many physical aspects. First of all, such a structure gives an interesting example of a nonintegrable system. Then, because of inserting a piece of a magnetized ferrite into the resonator domain, a microwave resonator behaves under odd time-reversal symmetry. A ferrite disk may act as a topological defect causing induced vortices. Such vortices can appear due to gyrotropy of a ferrite material. Finally, the studies of this object may clarify some important aspects concerning recent experiments of interaction of oscillating magnetic-dipolar modes (MDM) in ferrite disks with microwave electromagnetic fields.

Nonintegrable systems (such, for example, as Sinai billiards) are the subject for intensive studies in microwave experiments [1]. The results show that microwave cavity experiments can be an ideal laboratory for studying the so-called quantum-classical correspondence, a central issue in quantum chaos. To get microwave billiards with broken time-reversal symmetry, ferrite strips and ferrite cylinders were introduced into the resonators [2]. A role of material gyrotropy, as a factor provoking creation of vortices, was a subject of recent studies [3]. The vortices are defined as lines of powerflow, i.e. lines to which the Poynting vector is tangential. It is pointed out that in these experiments, the gyrotropic effect cannot be considered separately from losses. At the same time, due to just only inserted cylinders made of materials with losses (without any gyrotropic properties), vortices of the power flows can be observed in a rectangular billiard [4].

Most of the investigated nonintegrable systems are considered as hard-wall billiards. However, for the class of optical, or dielectric, nonintegrable systems the boundary manifests itself by a change in the index of refraction. The interplay of reflection and transmission at the different interfaces gives rise to a rich variety of wave phenomena [5]. The problem of non-hard-wall inclusions in microwave resonators suggests certain questions about internal fields inside these inclusions, especially when intrinsic material resonances (such, for example, as FMR or plasmon resonance) occur. In papers [2, 3], ferrite inclusions with the FMR conditions were used, but no proper analysis of the fields inside small ferrite samples were made. These "internal" fields may exhibit, however, very unique properties. In particular, recent studies [6] shows that for a case of a nanoparticle illuminated by the electromagnetic field, the "energy sink" vortices with spiral energy flow line trajectories are seen in the proximity of the nanoparticle's plasmon resonance.

The purpose of this paper is to study numerically the fields outside and inside a ferrite disk placed in a rectangular-waveguide cavity. This may call special attention in a view of a strong present interest in microwave nonintegrable systems with broken time-reversal symmetry. The disk is very thin compared to the waveguide height. Such a quasi two-dimensional nonintegrable system distinguished from configurations analyzed in [2, 3]. There is also a particular interest in such thin-film ferrite samples in a view of recent studies of interaction of MDM oscillations with cavity EM fields. Experimental studies show unique properties of interaction of oscillating MDMs in small ferrite disk resonators with microwave-cavity electromagnetic fields [7 − 10]. The character of the experimental multi-resonance absorption spectra leads to a clear conclusion that the energy of a source of a DC magnetic field is absorbing "by portions", or discretely, in other words [11]. The MDMs in a ferrite disk are characterized by dynamical symmetry breaking [12]. In this paper, we put aside any detailed discussions of MDM spectra in a disk, but study just how the FMR and geometry factors in a system (a microwave cavity plus a gyrotropic disk) may effect on the electromagnetic field patterns. It was assumed in [7 − 10], the cavity-field structure is not strongly perturbed by a ferrite sample. So it was supposed that the acting RF field corresponds to the original cavity field in a point where a ferrite sample is placed. As we will show in this paper, even a small FMR-disk may strongly perturb the cavity field. The power-flow lines of the microwave-cavity field interacting with a ferrite disk, in the proximity of its ferromagnetic resonance, may form whirlpool-like electromagnetic vortices. The role of ohmic losses in waveguide walls and dielectric and magnetic losses in a disk is a subject of our investigations. We study the symmetry properties of the vortices. There are certain symmetry features of vortices when one reverses the DC magnetic field and interchanges the RF source and receiver positions.

Since the nonintegrable nature of the problem precludes exact analytical results for the eigenvalues and eigenfunctions, numerical approaches are required. Numerical studies of vortex formations in different electromagnetic structures were a subject of numerous investigations. It concerns, for example, vortices of the Poynting vector field in the near zone of antennas [13], vortex formation near an iris in a rectangular waveguide [14], light transmission through a subwavelength slit [15], energy flow in photonic crystal waveguide [16]. Using the HFSS (the software based on FEM method produced by ANSOFT Company) and the CST MWS (the software based on FITD method produced by Computer Simulation Technology Company) CAD simulation programs for 3D numerical modeling of Maxwell equations, we are able to characterize the complete complex signal including the intensity of the signal and its phase relative to the incoming reference wave. In our numerical experiments, both modulus and phase of the fields are determined. It allows reconstructing the Poynting vector at any point within the resonator. The main results of numerical simulations we obtained based on the HFSS program since this program is more relevant for precise analyzing high-resonant microwave objects. The CST program was used just as a test program for some cases. All the below pictures correspond to the HFSS-program numerical results.

## 2. Cavity field structure and vortices

Let us consider a vacuum region of the cavity space. For TE polarized (with respect to the $y$-direction) electromagnetic waves, the singular features of the complex electric field component $E_y(x,z)$ can be related to those that will subsequently appear in the associated two-dimensional time-averaged real-valued Poynting vector field $\vec{S}(x,z)$.

The transport of electromagnetic energy through the resonator is described by the Poynting vector

$$\vec{S} = \frac{c}{4\pi}\left[\operatorname{Re}\left(\vec{E}_c e^{i\omega t}\right) \times \operatorname{Re}\left(\vec{H}_c e^{i\omega t}\right)\right], \tag{1}$$

where $\vec{E}_c$ and $\vec{H}_c$ are complex amplitudes of the field vectors. From the Maxwell equation in a vacuum one has

$$\operatorname{Re}\left(\vec{H}_c e^{i\omega t}\right) = \frac{c}{\omega}\operatorname{Im}\left[\nabla \times \left(\vec{E}_c e^{i\omega t}\right)\right]. \tag{2}$$

So Eq. (1) can be rewritten as

$$\vec{S} = \frac{c^2}{4\pi\omega}\left[\mathrm{Re}(\vec{E}_c e^{i\omega t}) \times (\nabla \times \mathrm{Im}(\vec{E}_c e^{i\omega t}))\right]. \tag{3}$$

We take advantage now of the following vector relation for two arbitrary vectors, $\vec{a}$ and $\vec{b}$. If one supposes that these vectors have only one component (let it will be the *y*-component), one evidently has: $\vec{a} \times (\nabla \times \vec{b}) = a_y \vec{\nabla}_\perp b_y$, where $\nabla_\perp$ is the differential operator with respect to the *x* and *z* coordinates. For electromagnetic fields, which are invariant with respect to the *y*-direction, this gives possibility to represent a time-average part of the Poynting vector as

$$\langle \vec{S} \rangle = \frac{c^2}{8\pi\omega} \mathrm{Im}(E_y^* \vec{\nabla}_\perp E_y), \tag{4}$$

where $E_y$ is a complex vector of the *y*-component of the electric field: $E_y \equiv (E_c)_y e^{i\omega t}$.

The fact that for electromagnetic fields invariant with respect to a certain coordinate, a time-average part of the Poynting vector can be approximated by a scalar wave function, allows analyzing the vortex phenomena. For a TE polarized field, we can write

$$E_y(x,z) \equiv \psi(x,z) = \rho(x,z) e^{i\chi(x,z)}, \tag{5}$$

where $\rho$ is an amplitude and $\chi$ is a phase of a scalar wave function $\psi$. One can rewrite Eq. (4) as

$$\langle \vec{S} \rangle = \rho(x,z)^2 \nabla_\perp \chi(x,z). \tag{6}$$

This representation of the Poynting vector in a quasi-two-dimensional system allows clearly define a phase singularity as a point (*x*, *z*) where the amplitude $\rho$ is zero and hence the phase $\chi$ is undefined. Such singular points of $E_y(x,z)$ correspond to vortices of the power flow $\langle \vec{S} \rangle$, around which the power flow circulates. A center is referred to as a (positive or negative) topological charge. Since such a center occur in free space without energy absorption, it is evident that $\nabla_\perp \cdot \langle \vec{S} \rangle = 0$. Because of the phase singularities of the free-space electromagnetic field, unique properties of the power flow transmission near a sub-wavelength slit can be demonstrated [15]. The fact that vortices of the free-space Poynting vector in flat electromagnetic resonators appear as a consequence of the nontrivial topological structure, make this relevant for modeling quantum vortices [1-4].

It was shown in [3] that due to the ferrite flat-boundary reflection effect, for TE polarized plain electromagnetic wave, there are microwave vortices of the Poynting vector in a vacuum region. For a waveguide structure, we reconstruct these results of the Poynting vector distribution based on the CAD simulation program. The HFSS-program results are shown in Fig.1 for a lossless [the perfect-electric-conductor (PEC) walls] rectangular waveguide with an enclosed lossless ferrite slab. A waveguide is terminated by the PEC wall. A system is exited by the $TE_{40}$ mode with the electric field oriented along $y$ axis. A ferrite [yttrium iron garnet (YIG)] is saturated ($4\pi M_s = 1880\,G$) by a DC magnetic field directed along $y$ axis. The working frequency (8.7 $GHz$) and a quantity of a bias magnetic field (5030 $Oe$) correspond to necessary conditions for a ferromagnetic resonance: a diagonal component of the permeability tensor is $\mu/\mu_0 = 23.85$ and an off-diagonal component is equal to $\mu_a/\mu_0 = 22.55$ [17]. In Fig. 1, one sees that in a vacuum region of the cavity space the singular points of $\langle \vec{S} \rangle$ (the vortex cores) can be directly related to the topological features of $E_y(x,z)$ (where the electric field is zero). It is also interesting to note that the vortex positions correspond to the regions with the homogeneous RF magnetic field.

Let us consider now a ferrite region of the cavity space. The transport of electromagnetic energy through this region is characterized by the Poynting vector described by Eq. (1). However, one cannot express now the Poynting vector just only by the $\vec{E}$-field vector, like it was shown in Eq. (3). So even if we suppose that there exists only one component of the electric field, $E_y(x,z)$, we cannot expect a priori to find any vortices of the power flow $\langle \vec{S} \rangle$ at singular points of $E_y(x,z)$. On the other hand, we can expect to find a vortex in a region of a maximal field $E_y(x,z)$.

## 3. The Poynting-vector vortices in a cavity with a ferrite disk

The main systems under our investigations are *X*-band rectangular-waveguide cavities with an enclosed thin ferrite disk. The systems resemble ones used in experiments [8-10]. There are a short-wall cavity with an iris [Fig. 2] and a cavity with two irises [Fig. 3]. The disk is very thin compared to the waveguide height and is placed in the middle of the waveguide height. The DC magnetic field ($H_0 = 5030\ Oe$) is normal to the disk plane. As a starting point of the study, we introduce in our CAD program the following material parameters of the disk: dielectric constant, $\varepsilon_r = 15$; dielectric losses, $\tan\delta = 0.01$; magnetic losses, $\Delta H = 0.1 Oe$. Metal walls of a cavity are made of copper and are characterized by the conductivity of $\sigma = 5.8 \times 10^7\ Siemens/m$. The microwave power enters to a cavity through an iris. The cavity operates at the $TE_{104}$ mode. By virtue of such an elongated structure one can clearly observe the power flow distribution in a system. The cavity resonances were estimated via frequency dependent absorption peaks of the $S_{11}$ and $S_{21}$ parameters of the scattering matrix.

A ferrite disk may act as a topological defect causing induced vortices. Figs. 4 (a), (b) and Figs. 5 (a), (b) show the streamlines of the Poynting vector in a short-wall cavity when a ferrite disk (diameter 6 mm, thickness 0.1 mm) is inserted in a maximum of the RF electric field ($l = \lambda/4$). One can distinguish the clockwise and counter-clockwise rotations of the power flow corresponding to two opposite orientations of a normal bias field. The vortex center – the topological singularity – is in a geometrical center of a disk. Fig. 6 shows the Poynting vector distribution immediately inside a ferrite disk. Figs. 7 and 8 show, respectively, the electric and magnetic RF fields in a disk. An analysis of the vortex pictures in Figs. 4 and Fig. 6 gives a result worth special notice. One can see that for the same direction of a bias magnetic field, there are opposite directions of Poynting-vector rotations inside and outside a ferrite disk.

A character of the vortex picture, in general, is independent from a disk diameter. This is illustrated in Fig. 9 which shows the Poynting vector distribution for a "big" disk (diameter 12 mm, thickness 0.1 mm). A vortex in a cavity with a ferrite disk takes a shape of a whirlpool. It becomes evident from a picture of the Poynting vector distribution immediately upper and below a ferrite disk. Such a picture is shown in Fig. 10 for a "big" disk. One can see the "energy-sink" character of the vortex with spiral energy flow line trajectories in the proximity of the FMR disk. A role of gyrotropy in forming the vortex structure becomes evident when one compares the above power flow characteristics with the power flow characteristics for a case of a simple lossy dielectric disk. Fig. 11 demonstrates distribution of the Poynting vector in a cavity with a lossy dielectric disk (diameter 6 mm, thickness 0.1 mm, dielectric constant, $\varepsilon_r = 15$; dielectric losses, $\tan\delta = 0.01$) inside a rectangular-waveguide cavity.

The vortex pictures become essentially different when one places a ferrite disk in a maximum of the short-wall-cavity magnetic field ($l = \lambda/2$). These pictures, for a "small" disk (diameter 6 mm, thickness 0.1 mm) are shown in Figs. 12 and 13. The pictures correspond to two opposite orientations of a normal bias field. In this situation one can distinguish two coupled vortices having the same "topological charge" (the same direction of rotation for a given direction of the bias magnetic field). The vortices centers are shifted from a geometrical center of a disk and are situated near the disk border.

A change of the microwave source and receiver positions (at the same direction of the bias magnetic field) does not change the vortex rotation direction. Fig. 14 (a) shows the Poynting vector distribution in a two-irise cavity when a "big" (diameter 12 mm, thickness 0.1 mm) ferrite disk is placed in maximum of the electric field. The power input is at the left-hand side of a system. If one interchanges the microwave source and receiver positions, leaving fixed a direction of the bias magnetic field, the vortex will have the same rotation direction. This is shown in Fig. 14 (b). It means that the vortex rotation direction is invariant with respect to mirror reflection along a waveguide axis. When one reverses the DC magnetic field together with an interchange of the microwave source and receiver positions, the vortex changes its rotation direction [compare Figs. 14

(b) and 14 (c)]. It means that the vortex rotation direction is not invariant with respect to a combined symmetry operation: mirror reflection and time reversal.

### 4. The role of material losses in forming vortices

One of the main points in our studies should concern a role of the material losses in a system in forming the ferrite-disk vortex structures. An essential role of the material losses in forming the electromagnetic vortices was stressed in paper [6], where the spiral energy flow line trajectories in the proximity of the nanoparticle's plasmon resonance were studied for different quantities of the imaginary parts of the permittivity. We found that also in our case the vortex structure and the topography of the field maps depend on the values of dissipation losses.

For a case when a ferrite disk is placed in a maximum of the short-wall cavity electric field, we found that in the regions distant from a ferrite disk there is a strong vortex-picture dependence on the waveguide wall ohmic losses and the disk material losses. At the same time, the vortex structure inside a ferrite disk and in immediate proximity to a ferrite disk remains practically the same when one changes the wall conductivity and the disk losses. Fig. 15 (a) shows the Poynting-vector distribution in a short-wall cavity for the PEC waveguide walls and disk losses parameters: $\tan\delta = 0.01$, $\Delta H = 0.1 Oe$. In comparison with Figs. 4 and 5, one can see the difference in the Poynting-vector distributions in the cavity regions distant from a ferrite disk. Now, it appears an additional vortex in a vacuum region of the cavity. A character of this additional vortex changes when one changes the disk losses parameters. Fig. 15 (b) corresponds to parameters: PEC waveguide walls, $\Delta H = 0.1 Oe$, $\tan\delta = 0.0002$ and Fig. 15 (c) is for PEC walls, $\Delta H = 5 Oe$, $\tan\delta = 0.01$. Further increase of magnetic losses parameter $\Delta H$ leads to elimination of an additional vortex. Regarding the vortex pictures shown in Figs. 15 (a), (b), and (c) it is important to note that the cores of additional vortices are situated at a distance $l = \lambda/2$ from a short wall along z axis, evidently corresponding to regions where $E_y(x, z) = 0$.

## 6. Discussion and conclusion

From a graphical representation of the Poynting vector one sees that the absorption cross-section of a ferrite particle can be much bigger than its geometrical cross-section. When a vortex is created, power flow lines passing through the ferrite particle generate the high energy losses associated with the large microwave cross-section. The reason why a ferrite sample absorbs much more radiation than that given by the geometrical cross-section can be explained by the fact that the field enters into the particle not only from the front (with respect to a power source) part of the surface, but also from the back ("shadow") side. In other words, one can suppose that a disk absorbs incident energy through its whole surface. The feature of the observed Poynting-vector vortex is the fact that this vortex cannot be characterized by some invariant, such as the flux of vorticity. So a vorticity thread may not be defined for such a vortex.

The problem of microwave vortices of the Poynting vector created by ferrite disks looks as very important for many modern applications, e.g., for near-field microwave lenses, for field concentration in patterned microwave metamaterials, for high-Q resonance microwave devices. It is supposed that these "swirling" entities can, in principle, be used to carry data and point to new communication systems. Application of such generic ideas to microwave systems is of increasing importance in numerous utilizations. For example, any work in the area of a "vortex antenna" is likely to generate many unique microwave systems, e.g., in direction finding and target ranging. With respect to this aspect, an interesting phenomenon of the transition of a near-zone phase singularity into a singularity of the radiation pattern has been shown recently [18].

**Figure captions:**

Fig. 1. The $TE_{40}$ lossless waveguide structure with an enclosed lossless ferrite slab (crosses show the vortex positions).

(a) Distribution of the Poynting vector;

(b) Distribution of the electric field.

Fig. 2. A short-wall cavity with an iris and an enclosed ferrite disk.

Fig. 3. A cavity with two irises and an enclosed ferrite disk.

Fig. 4. The cavity Poynting vector distribution. A ferrite disk is placed in the maximum of the cavity electric field. A bias field is directed along positive y axis. (a) A general view; (b) An enlarged picture of the vortex.

Fig. 5. The cavity Poynting vector distribution. A ferrite disk is placed in the maximum of the cavity electric field. A bias field is directed along negative y axis.

(a) A general view;

(b) An enlarged picture of the vortex.

Fig. 6. The Poynting vector distribution inside a ferrite disk.

Fig. 7. Electric field vector inside a ferrite disk.

Fig. 8. Magnetic field vector inside a ferrite disk.

Fig. 9. A vortex created by a "big" ferrite disk.

(a) A general view;

(b) An enlarged picture of the vortex.

Fig. 10. Illustration of the whirlpool-like character of the Poynting-vector vortex.

Fig. 11. The Poynting vector distribution with a lossy dielectric disk in a maximum of the cavity electric field.

Fig. 12. The cavity Poynting vector distribution. A ferrite disk is placed in the maximum of the cavity magnetic field. A bias field is directed along positive y axis.

Fig. 13. The cavity Poynting vector distribution. A ferrite disk is placed in the maximum of the cavity magnetic field. A bias field is directed along negative *y* axis.

Fig. 14. The Poynting-vector distribution in a two-iris cavity with a "big" ferrite disk in a maximum of the electric field.

(a) The input is at the left-hand side of a system. $\vec{H}_0$ in the positive y direction.

(b) The input is at the right-hand side of a system. $\vec{H}_0$ in the positive y direction.

(c) The input is at the left-hand side of a system. $\vec{H}_0$ in the negative y direction.

Fig. 15. The Poynting-vector distribution for the PEC waveguide walls and different disk losses parameters.

(a) $\Delta H = 0.1 Oe, \tan\delta = 0.01$; (b) $\Delta H = 0.1 Oe, \tan\delta = 0.0002$; (c) $\Delta H = 5 Oe, \tan\delta = 0.01$

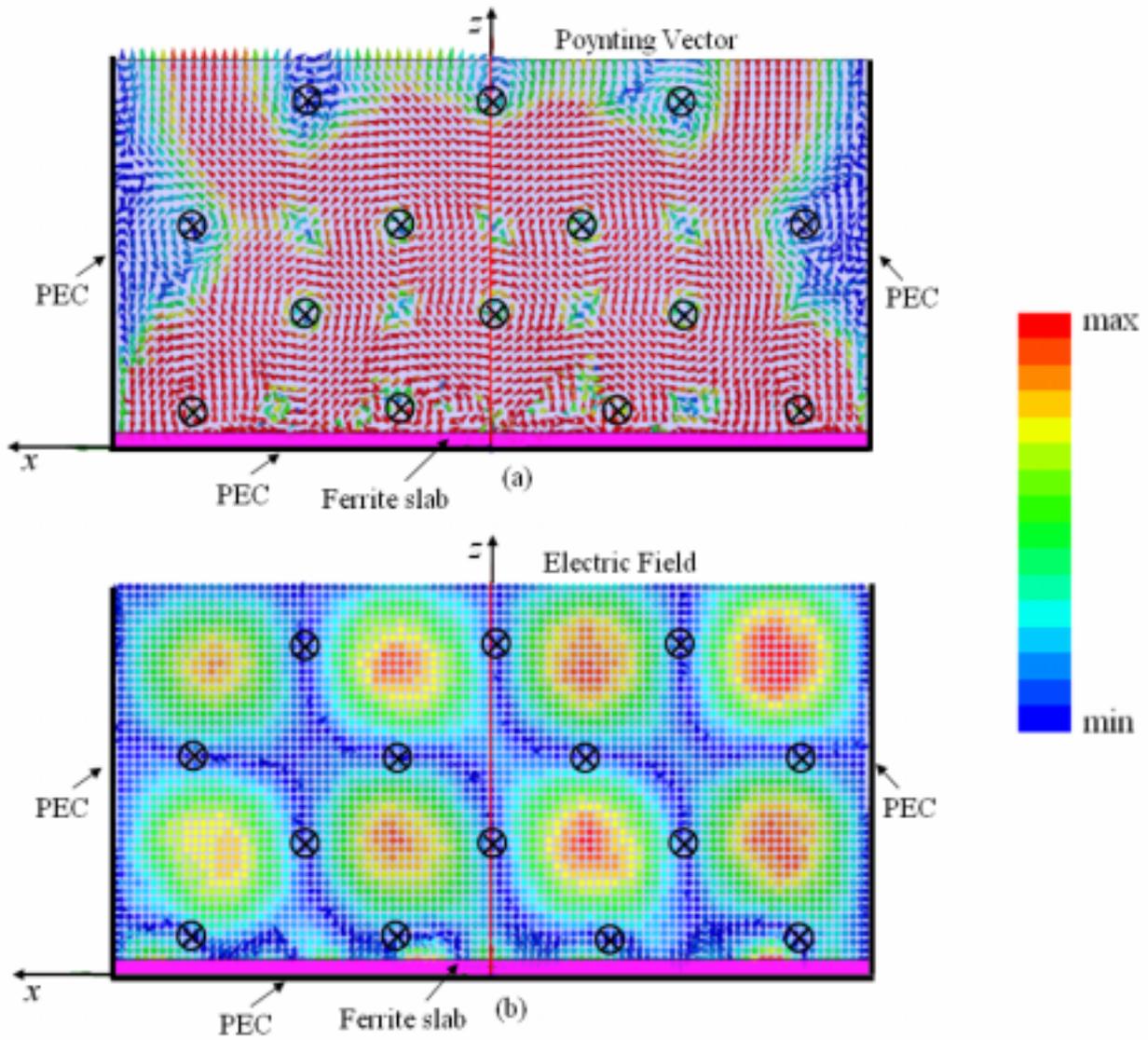

Fig. 1. The TE$_{40}$ lossless waveguide structure with an enclosed lossless ferrite slab (crosses show the vortex positions).

(a) Distribution of the Poynting vector

(b) Distribution of the electric field.

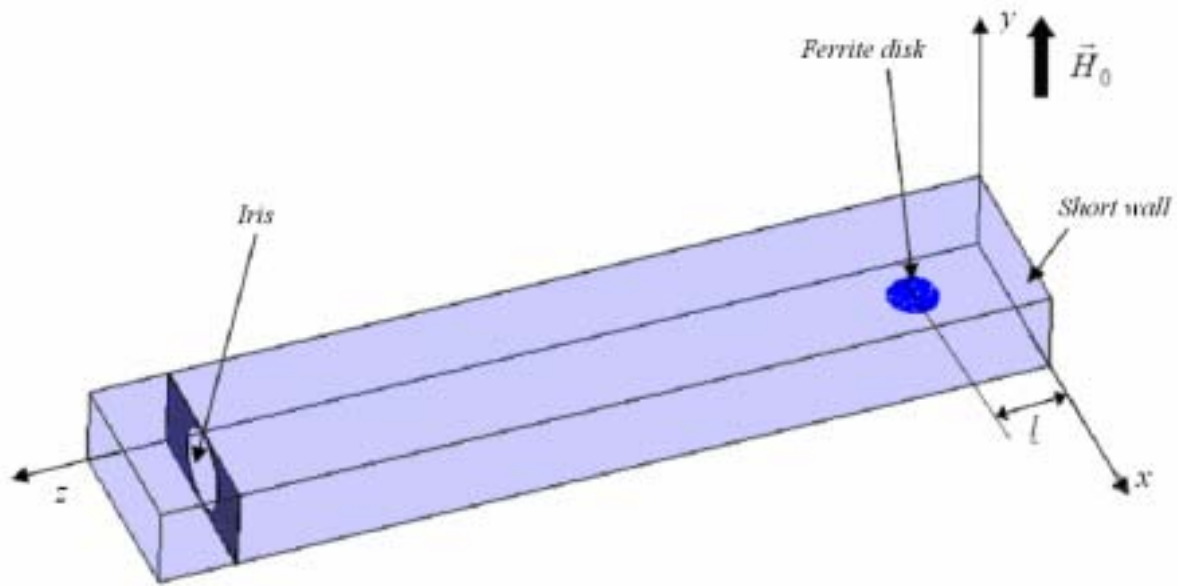

Fig. 2. A short-wall cavity with an iris and an enclosed ferrite disk

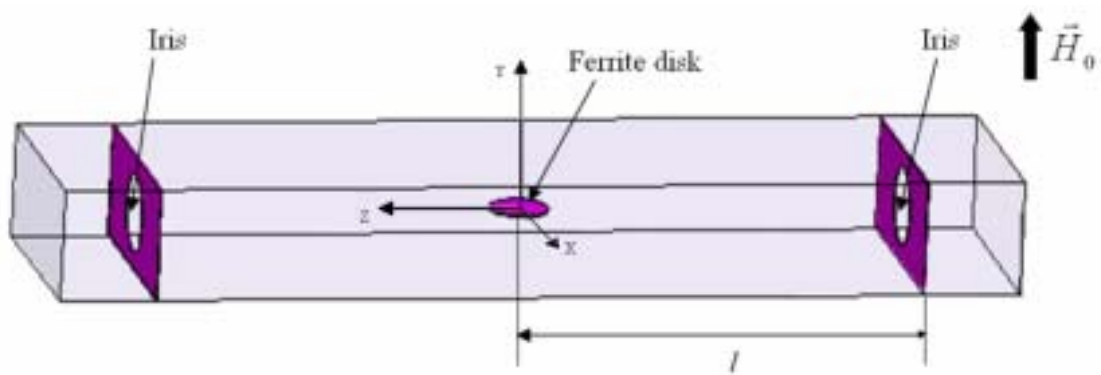

Fig. 3. A cavity with two irises and an enclosed ferrite disk.

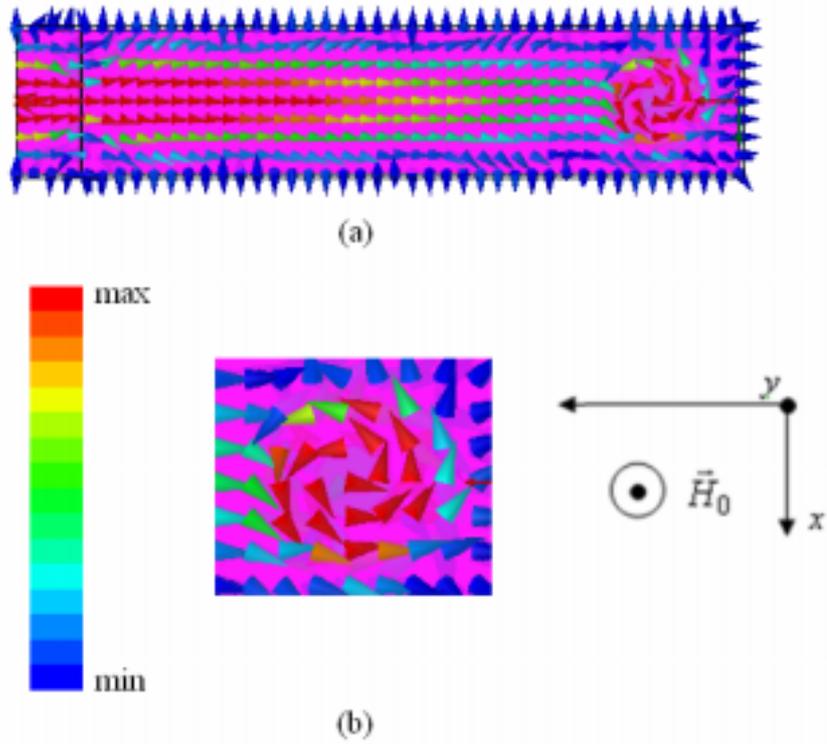

Fig.4. The cavity Poynting vector distribution. A ferrite disk is placed in the maximum of the cavity electric field. A bias field is directed along positive *y* axis.
  (a) A general view;
  (b) An enlarged picture of the vortex.

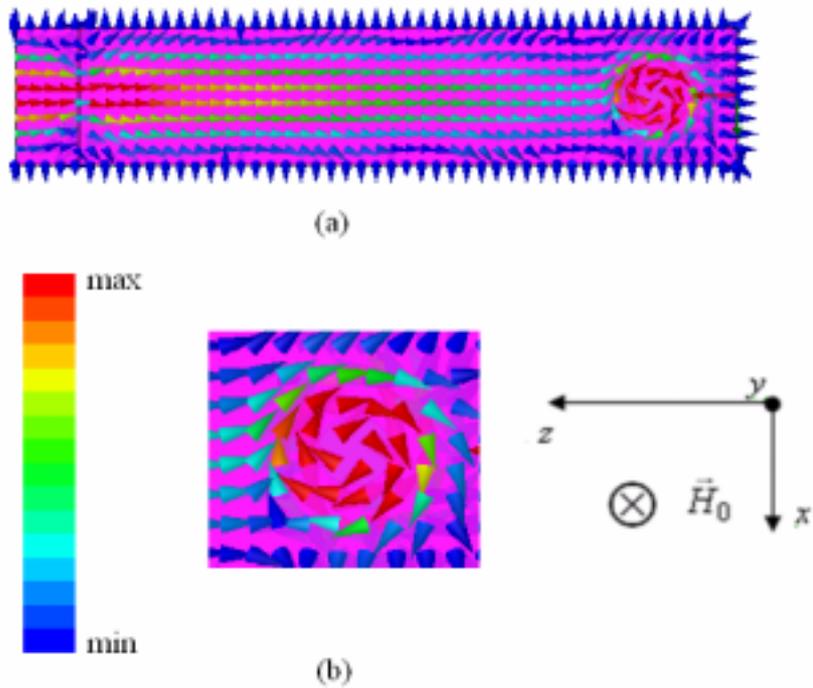

Fig.5. The cavity Poynting vector distribution. A ferrite disk is placed in the maximum of the cavity electric field. A bias field is directed along negative *y* axis.
  (a) A general view;
  (b) An enlarged picture of the vortex.

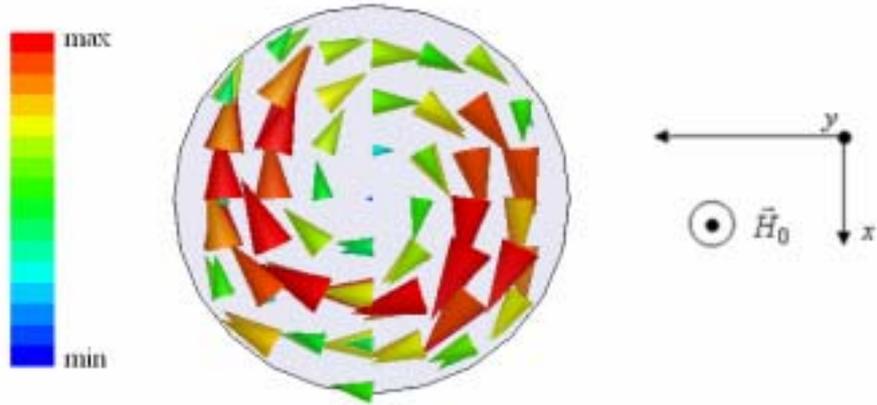

Fig. 6. The Poynting vector distribution inside a ferrite disk.

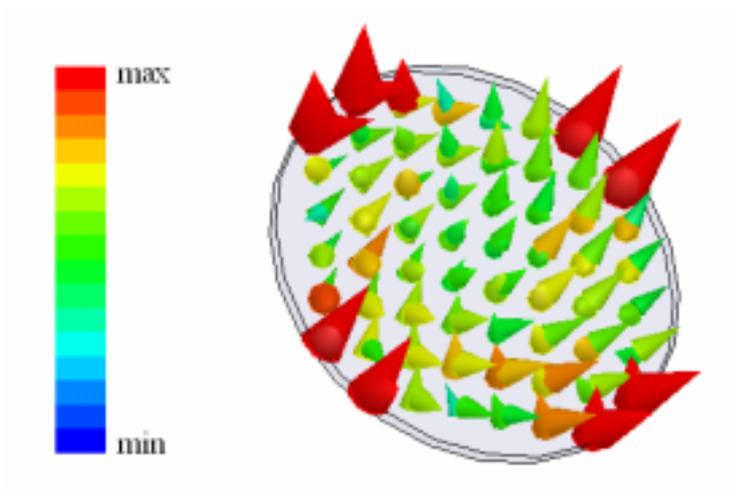

Fig. 7. Electric field vector inside a ferrite disk.

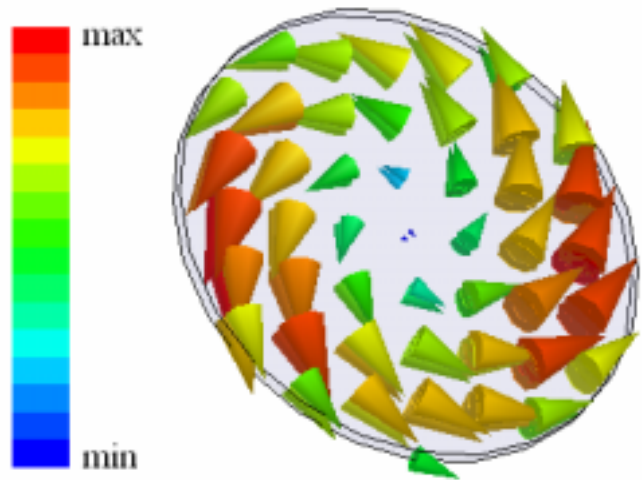

Fig. 8. Magnetic field vector inside a ferrite disk.

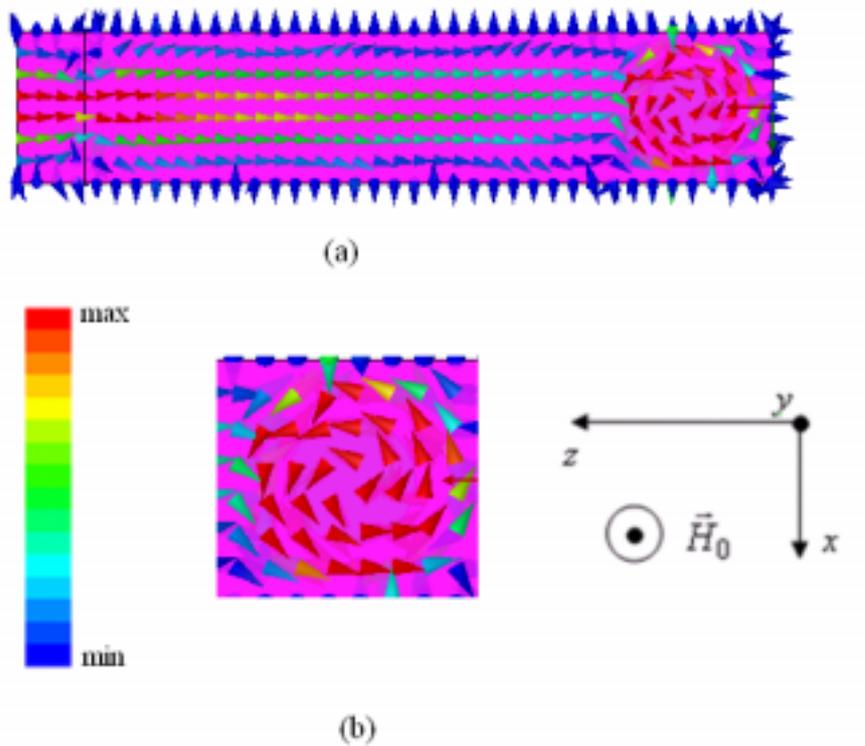

Fig. 9. A vortex created by a "big" ferrite disk.

(a) A general view;

(b) An enlarged picture of the vortex.

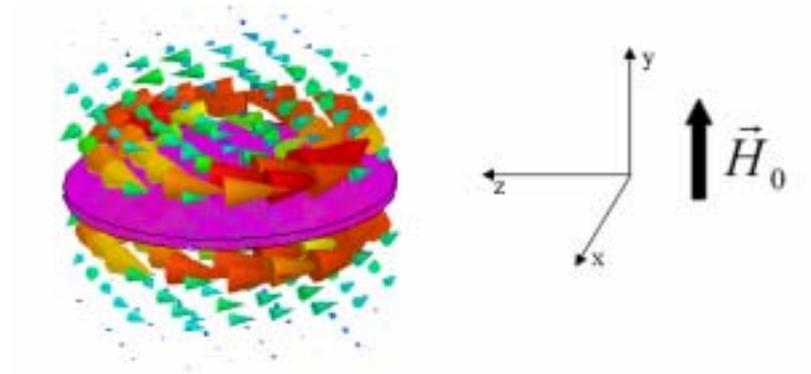

Fig. 10. Illustration of the whirlpool-like character of the Poynting-vector vortex.

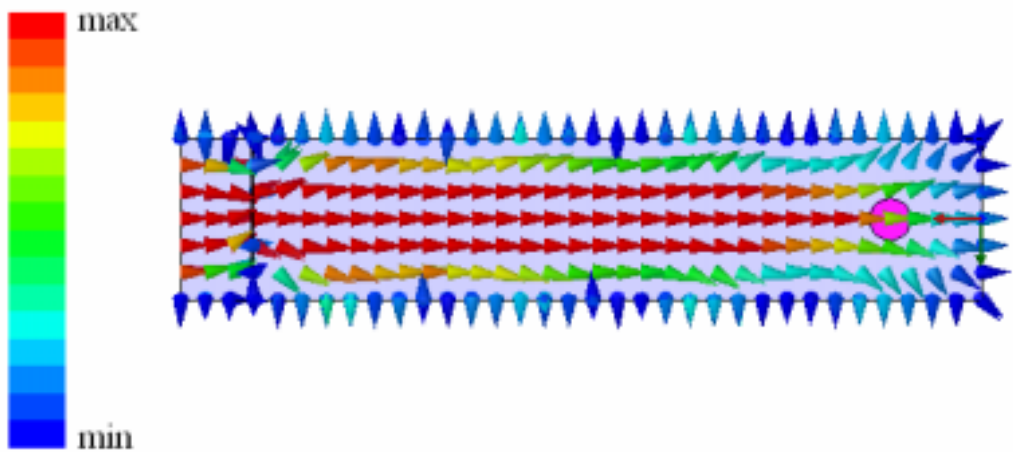

Fig. 11. The Poynting vector distribution with a lossy dielectric disk in a maximum of the cavity electric field.

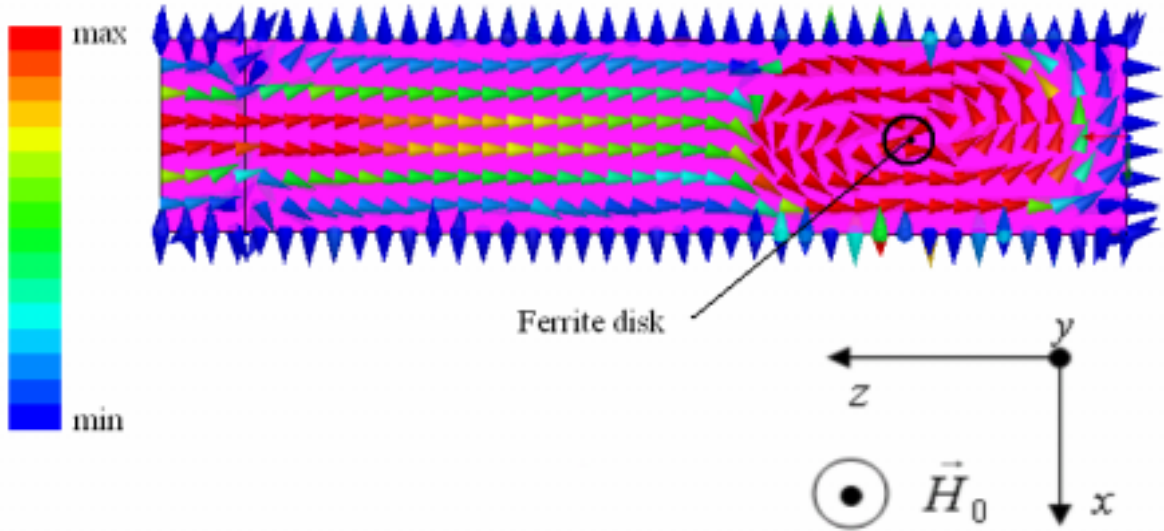

Fig.12. The cavity Poynting vector distribution. A ferrite disk is placed in the maximum of the cavity magnetic field. A bias field is directed along positive *y* axis.

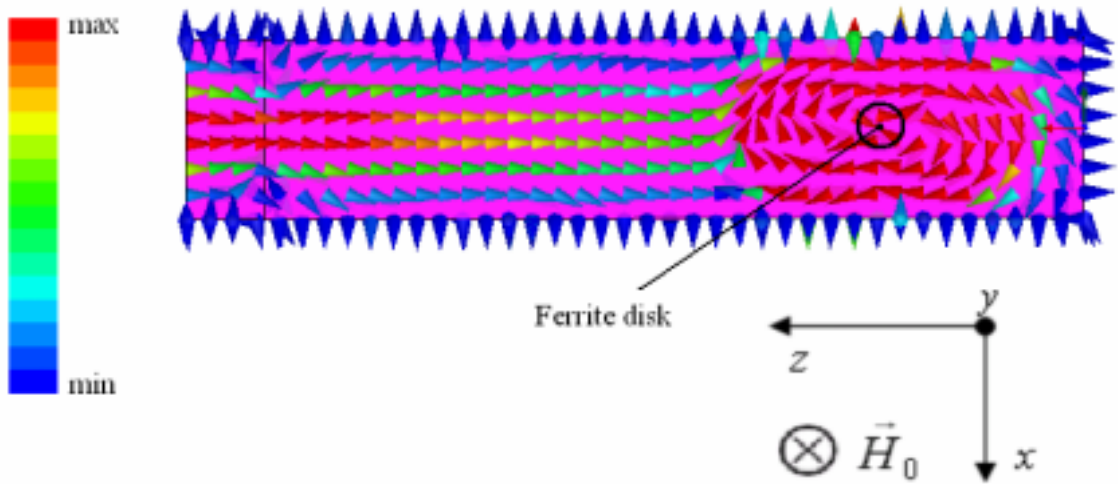

Fig.13. The cavity Poynting vector distribution. A ferrite disk is placed in the maximum of the cavity magnetic field. A bias field is directed along negative *y* axis.

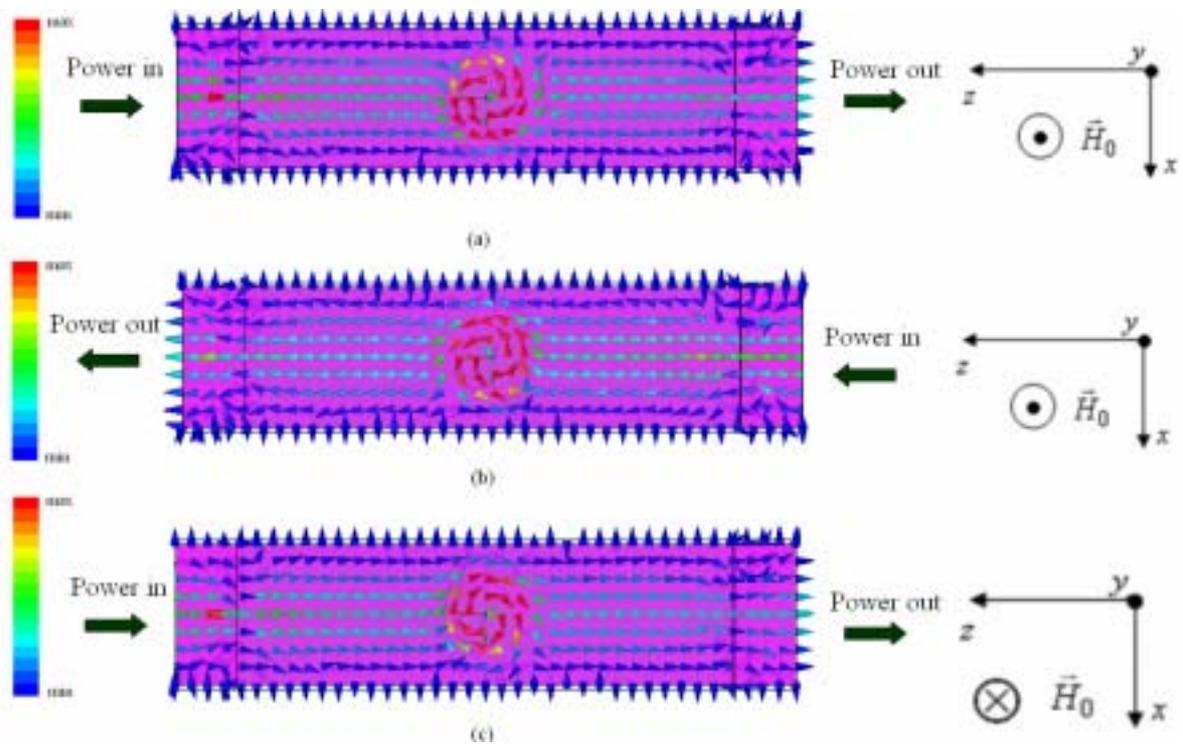

Fig. 14. The Poynting-vector distribution in a two-iris cavity with a "big" ferrite disk in a maximum of the electric field.

a) The input is at the left-hand side of a system. $\vec{H}_0$ in the positive *y* direction.

b) The input is at the right-hand side of a system. $\vec{H}_0$ in the positive *y* direction.

c) The input is at the left-hand side of a system. $\vec{H}_0$ in the negative *y* direction.

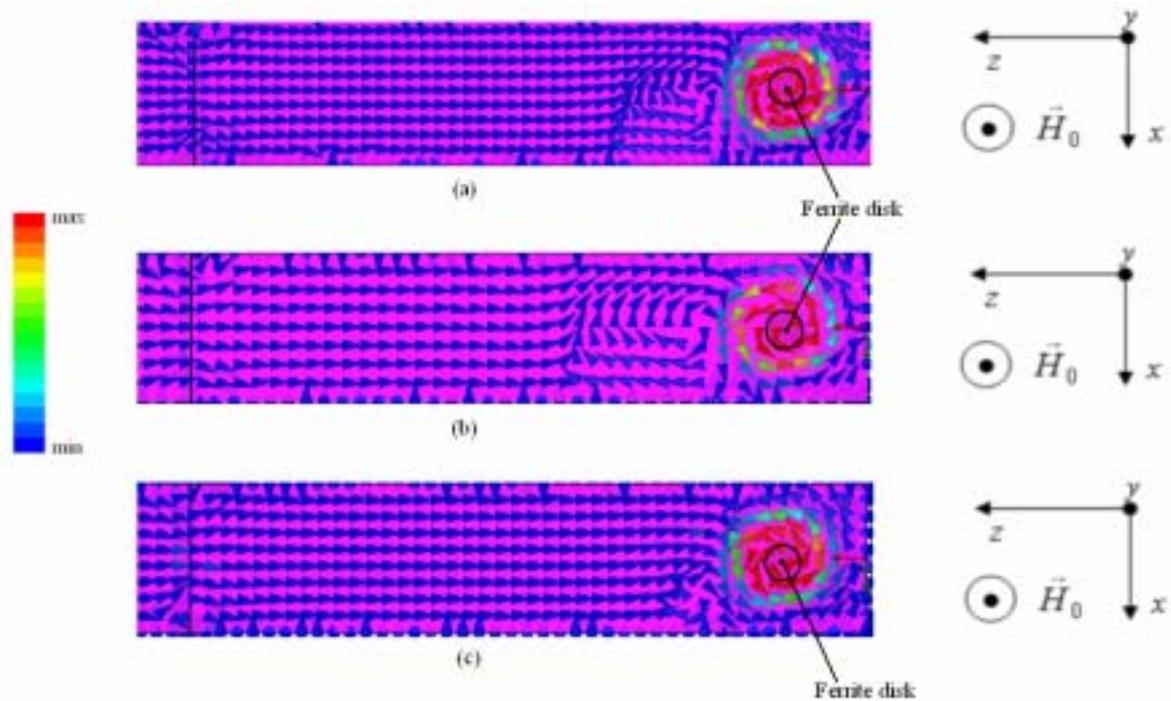

Fig. 15. The Poynting-vector distribution for the PEC waveguide walls and different disk losses parameters.

(a) $\Delta H = 0.1 Oe, \ \tan\delta = 0.01$;

(b) $\Delta H = 0.1 Oe, \ \tan\delta = 0.0002$;

(c) $\Delta H = 5 Oe, \ \tan\delta = 0.01$